\documentclass[prl, aps, preprint, longbibliography, superscriptaddress]{revtex4-2}
\usepackage[pretty, uselistings]{revquantum}
\usepackage{graphicx}
\usepackage{dcolumn}
\usepackage{bm}
\usepackage{threeparttable}
\usepackage[T1]{fontenc}
\usepackage[utf8]{inputenc}
\usepackage{gensymb}
\usepackage{amsmath} 
\usepackage{amsfonts} 
\usepackage{color}
\usepackage{natbib}
\usepackage{longtable}
\begin{document}
	\title{Deterministic coupling of a quantum emitter to  surface plasmon polaritons,  Purcell enhanced generation of indistinguishable single photons and quantum information processing}
	\author{Lakshminarayan Sharma}	
	\affiliation{Department of Physics, Birla Institute of Technology, Mesra, Ranchi-835215, Jharkhand, India}
	\author{Laxmi Narayan Tripathi*}
	\affiliation{Department of Physics, School of Advanced Sceinces, Vellore Institute of Technology, Vellore-632014, Tamilnadu, India}
	\email{nara.laxmi@gmail.com}
	\date{\today}
\begin{abstract}
	Integrated photonic circuits are an integral part of all-optical and on-chip quantum information processing and quantum computer. Deterministically integrated single-photon sources in nanoplasmonic circuits lead to densely packed scalable quantum logic circuits operating beyond the diffraction limit. Here, we report the coupling efficiency of single-photon sources to the plasmonic waveguide, characteristic transmission spectrum, propagation length, decay length, and plasmonic Purcell factor.  We simulated the transmission spectrum to find the appropriate wavelength for various width of the dielectric in the metal-dielectric-metal waveguide. We find the maximum propagation length of 3.98 $\mu$m for Al$_{2}$O$_{3}$   dielectric-width equal to 140 nm and coupling efficiency to be greater than  82 \%. The plasmonic Purcell factor was found to be inversely proportional to dielectric-width (w), reaching as high as 31974 for w equal to 1 nm. We also calculated quantum properties of the photons like indistinguishability and found that it can be enhanced by plasmonic-nanocavity if single-photon sources are deterministically coupled. We further, propose a scalable metal-dielectric-metal waveguide based quantum logic circuits using the plasmonic circuit and Mach-Zehnder interferometer.
\end{abstract}


\maketitle

\section{Introduction}

Surface plasmon polaritons (SPP) are electromagnetic surface waves coupled to free electron oscillations in metal. They propagate along the metal-dielectric boundary with the transverse amplitude decaying exponentially at both sides \cite{Yang2017, Han2014c, Bozhevolnyi2006a}. The smallest dimension d, to which light can be concentrated is limited by the diffraction, which is d $\geq$ $\lambda$/2n  where  $\lambda$ is the wavelength of the electromagnetic wave, n is the refractive index of the medium. SPPs break the classical diffraction limit  \cite{Maier2007} and can confine light to a nanoscale length regime which makes them an attractive plasmonic platform for on-chip integrated quantum photonic circuits \cite{Babicheva2013, Boltasseva2005, Dionne2005, Ozbay2006, Zia2006, Zia2004}. A diverse type of geometrical structures have been used for SPPs propagation like metal-dielectric-metal (MDM) wave-guides \cite{Fang2015, Matsuzaki2008, Bian2014, Kurokawa2007, Caligiuri2019, Khurgin2012,Jun2008,Faggiani2015, Sondergaard2008},  slot wave-guides \cite{Veronis2005, Dionne2006, Dionne2006a} and channel plasmon waveguide \cite{Bozhevolnyi2006}. With the photonic integrated circuits several numbers of complex optical and electronic circuits can be miniaturized on a single chip \cite{Zia2006, Cai2010, Chandran2012, Barnes2003}. Linear optical quantum information processing requires single-photon sources (SPS) \cite{Knill2001}. Several SPSs have been identified in the past decades such as quantum dots \cite{He2013d}, transition metal dichalcogenides (TMDC) \cite{Tripathi2018a}, hBN (hexagonal Boron Nitride) \cite{ Sajid2020, Grosso2017a}, Nitrogen vacancy in diamonds \cite{Andersen2017}, spontaneous parametric down-conversion in nonlinear crystals etc \cite{Eisaman2011}. Recently, it has been found that the coupling of single photons sources with cavity results in Purcell enhanced indistinguishability \cite{Saxena2019,  He2013e, Ding2016a}. 

Metal-dielectric-metal sub-wavelength plasmonic waveguides have been shown as beam splitters and bend \cite{Veronis2005a, Veronis2007a}, directional coupler, and Mach-Zehnder interferometer \cite{Pu2010}. As opposed to classical information processing using bits, which is either 0 or 1, quantum information processing involves superposition states \cite{OBrien2009, Vogl2019}.  Superposition states imply the probilististic nature of bits, 0 and 1 (known as quantum bits or qubits). Any measurement for 0 and 1 destroys the superposition state known as the no-cloning theorem \cite{Wootters1982}. This no-cloning property of the quantum states (superposition states) makes the communication secure.  The qubits or 0 and 1 can be realized in single photons by assigning 0  and 1 to orthogonally polarized states usually referred to as polarization qubits \cite{Krenn2016}. In this case, the horizontal polarization state can be ascribed to 0 and vertical polarization as 1. The polarization qubits of single photons can easily be generated, controlled, and further manipulated by common linear optical devices like waveplates.  The single photons thus offer the most prominent source of qubits. A polarization qubit is a coherent superposition of horizontal and vertical polarization-maintaining a certain phase relation among them. Such a qubit is equivalent to a photon polarized at + 45$^{\circ}$. Setting a polarizer at the angle + 45$^{\circ}$ will allow the passing of the photon with 100 \% probability and with zero probability at - 45$^{\circ}$ \cite{Krenn2016} and thus linear optics is very well suited for quantum computation \cite{Rohde2015, Carolan2015}. 	
In this paper, we have investigated the SPP properties such as propagation length, decay length, and the plasmonic Purcell Factor change in Ag-Al$_{2}$O$_{3}$-Ag waveguide using Finite Difference Time Domain (FDTD) simulations and analytical calculations. We also calculated Purcell enhanced Indistinguishability of photons for various emitters coupled with a nano-plasmonic cavity. Our integrated  SPS and metal-dielectric-metal waveguide system are suitable for scalable plasmonic platform based high-density optical quantum processing circuits. We propose that using linear optical elements, our deterministically placed nanoantenna - coupled single-photon sources and Mach-Zehnder interferometer, a controlled-NOT gate can be formed.  

\section{ Finite difference time domain simulation design and theoretical model}

\begin{figure} 
	\centering
	\includegraphics[width=0.7\linewidth]{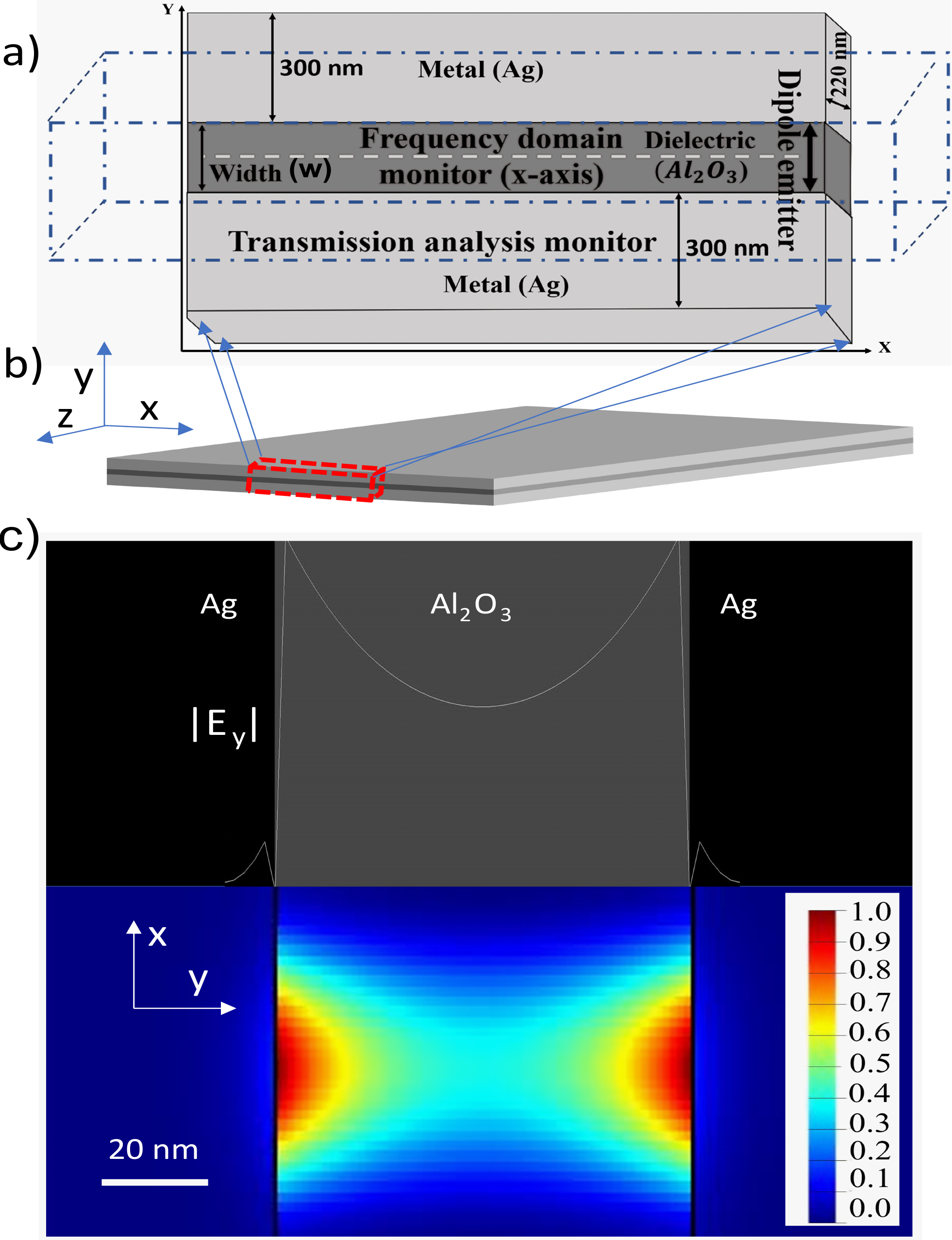}
	\caption{(a) FDTD simulation design of metal-dielectric-metal plasmonic waveguide with dipole emitter (in XY plane) embedded in the Al$ _{2} $O$ _{3}$ layer. (b) Thin film based easy to fabricate metal-dielectric-metal  waveguide. (c) Top: Transverse Ey-field profile in the waveguide for dielectric-width w equal to 80 nm at $\lambda$ equal to 437 nm.  Bottom:  XY cross section of TM1 mode (symmetrical mode) of the MDM waveguide. Color scale: Normalized magnitude of Ey-field.}  
	\label{fig:1}
\end{figure}
Our FDTD  simulation  MDM waveguide design  (Fig.~\ref{fig:1}) consists of two identical Ag layer of width 300 nm each  separated by the dielectric  (Al$_{2}$O$_{3}$) spacer of width w. The horizontal length is equal to 63 $\mu$m and the z span is 220 nm.  Transmission analysis monitor	(dashed cuboid box)  was used to  find the  wavelength 
of  maximum transmission ($\lambda_{max}$). The  frequency domain monitor (dashed white line passing through the dielectric) was kept along x-axis for measuring the field intensity. The structure was designed and modeled in Lumerical solutions FDTD and MODE software with  perfectly matched boundary keeping the mesh size equal to 5 nm. The transmission spectrum for the various dielectric-width ranging from 20 nm to 150 nm in step of 10 nm  was recorded to find the wavelength of maximum transmission ($\lambda_{max}$). The dielectric function of metals as described by Drude model \cite{Johnson1972} is 
\begin{equation}
\epsilon(m) = 1- \frac{\omega_{p}^{2}}{(\omega^{2}+\gamma\omega i)}
\label{eq1}
\end{equation}
here $\omega_{p}$ is the plasma frequency and $\gamma$ is the characteristic collision frequency.
The plasma frequency is given as \cite{Maier2007} 
\begin{equation}
\omega_{p} =  \frac{ne^{2}}{\epsilon_{o} m}
\label{eq2}
\end{equation}
where n is density of free electrons, e is electronic charge, $\epsilon_{o}$ is the permittivity of free space and m is mass of an electron. The value of $\omega_{p}$ the plasma frequency and $\gamma$ characteristic collision frequency used for Ag are $ 1.37  \times 10^4  $ THz  \cite{Herrera2014} and 100 THz  \cite{Maier2007} respectively and  $\epsilon_{d}$ for Al$_{2}$O$_{3}$  is 9 \cite{Robertson2004}.
\section{ Analytical calculation of surface plasmon polariton characteristics} 

For single interface between metal and dielectric, the dispersion relation of SPP is given by \cite{Maier2007} 
\begin{figure}
	\includegraphics[width=0.8\linewidth]{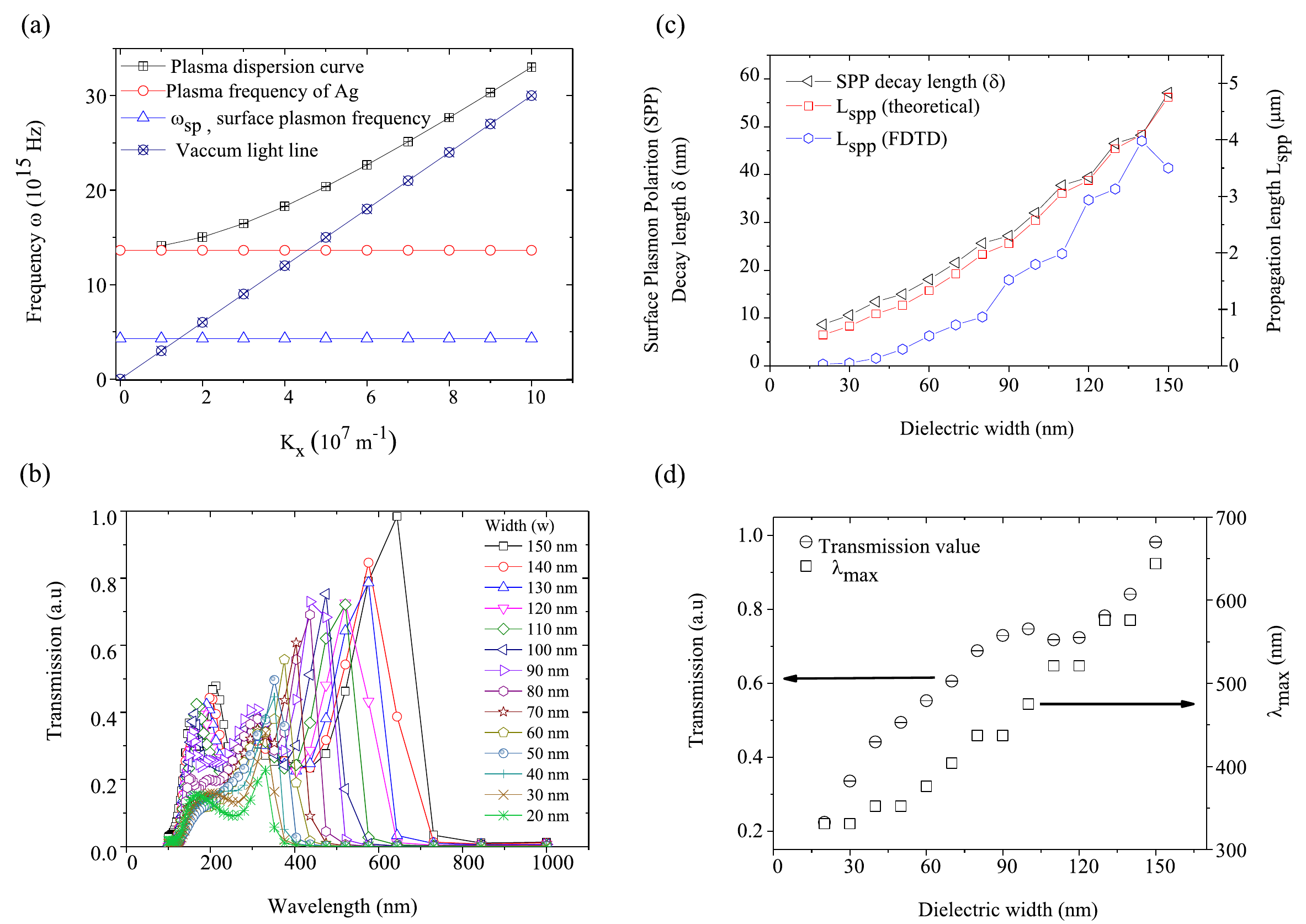}
	\caption{ (a) The plasma dispersion curve for silver (Ag), plasma frequency for Ag from equation~\ref{eq2} and the surface plasmon frequency for single interface with Al$_{2}$O$_{3}$ as the dielectric from equation~\ref{eq4} and the $\omega$ vs K relation for light in vacuum. (b) Transmission  vs
		wavelength plot for the various  dielectric-width (20 nm - 150 nm). (c) Surface plasmon polariton decay length ($\delta$), calculated surface plasmon polariton propagation length (L$_{spp}$) and the simulated propagation length (L$_{spp}$) from FDTD with increasing width of the dielectric layer from 20 nm to 150 nm in steps of 10 nm. (d) The peak transmission wavelength ($\lambda_{max}$) and the normalized transmission
		value plot for changing dielectric-width (w) of the MDM waveguide.}	
	\label{fig2}
\end{figure}

\begin{equation}
K_{SPP} = K_{o}\sqrt{\frac{\epsilon_{m}\epsilon_{d}}{\epsilon_{m}+\epsilon_{d}}}
\label{eq3}
\end{equation}
with $\epsilon_{m}$ and $\epsilon_{d}$ equal to dielectric permittivity of metal and dielectric respectively.

For single interface between metal and dielectric the large SPP wave vector value is approached when $\omega$ is equal to characteristic surface plasmon frequency $\omega_{sp}$ \cite{Maier2007}
\begin{equation}
\omega_{sp}=\frac{\omega_{p}}{\sqrt{1+\epsilon_{d}}}
\label{eq4}
\end{equation} 
$\omega_{sp}$ for Ag and  Al$_{2}$O$_{3}$ interface is $ 4.3 \times 10^3  $ THz.
\begin{figure*} 
	\centering
	\includegraphics[width=0.7\linewidth]{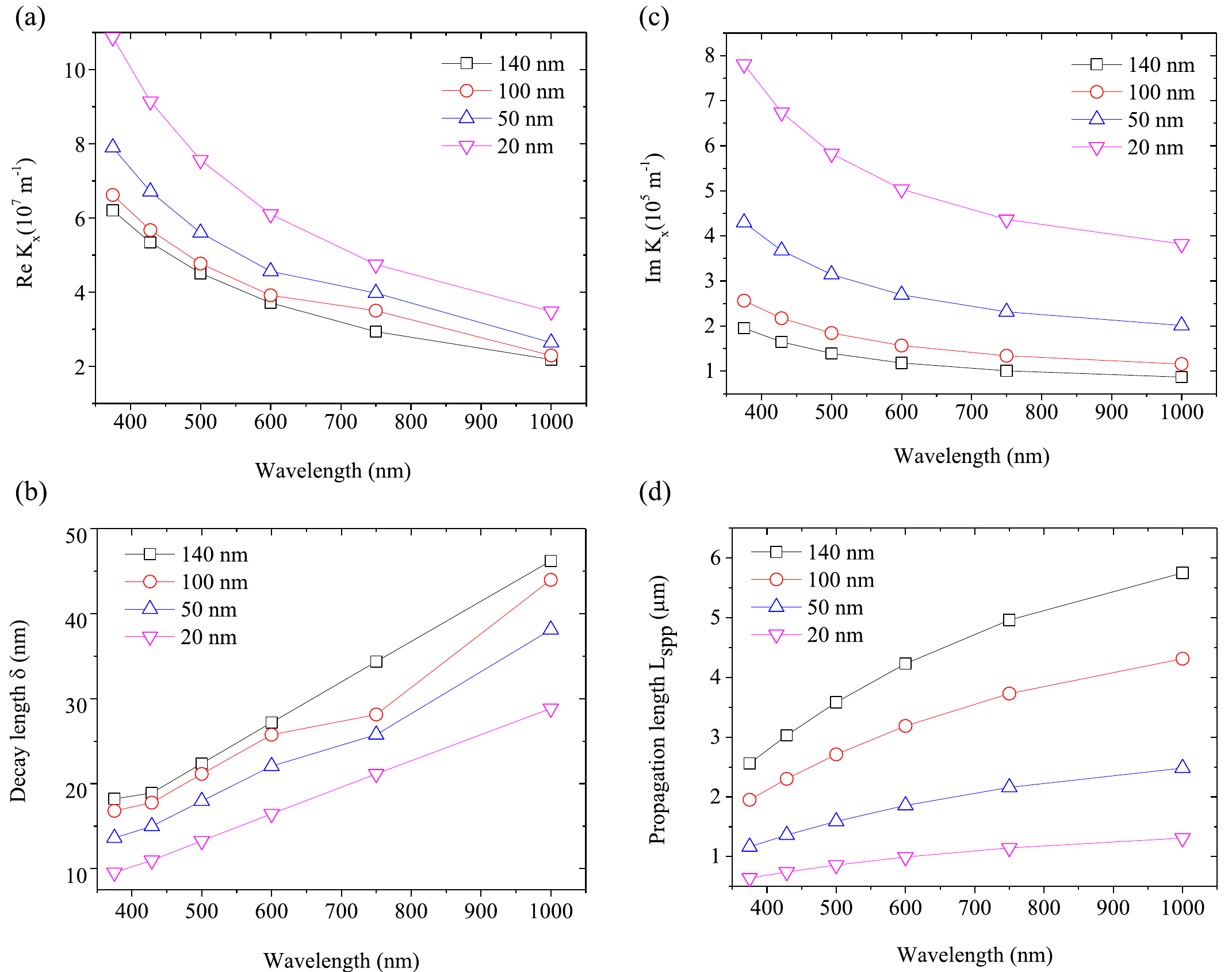}
	\caption{Surface plasmon polariton decay length ($\delta$) and propagation length (L$_{spp}$) variation with dipole emitter wavelength for various   dielectric-width (w) of the metal-dielectric-metal  wave-guide.  (a) Real part of K$_{x}$ vs   dipole emitter wavelength, (b) Surface plasmon polariton decay length ($\delta$) vs dipole emitter wavelength, (c) Imaginary part of K$_{_x}$ vs dipole emitter wavelength and (d) Surface plasmon polariton propagation length (L$_{spp}$) vs dipole emitter wavelength.}
	\label{fig3}
\end{figure*}

The complex propagation constant for SPP in multilayer system as MDM is given by \cite{Collin2007} 
\begin{equation}
K_{SPP} = n_{eff} K_{o}
\label{eq5}
\end{equation}
with
\begin{equation}
n_{eff}= (\epsilon_{d})^{1/2}\sqrt{1+\frac{\lambda}{\pi w (-\epsilon_{m})^{1/2}}(1+\frac{\epsilon_{d}}{-\epsilon_{m}})^{1/2}}
\label{eq6}
\end{equation}
\begin{figure*}
	\includegraphics[width=1\linewidth]{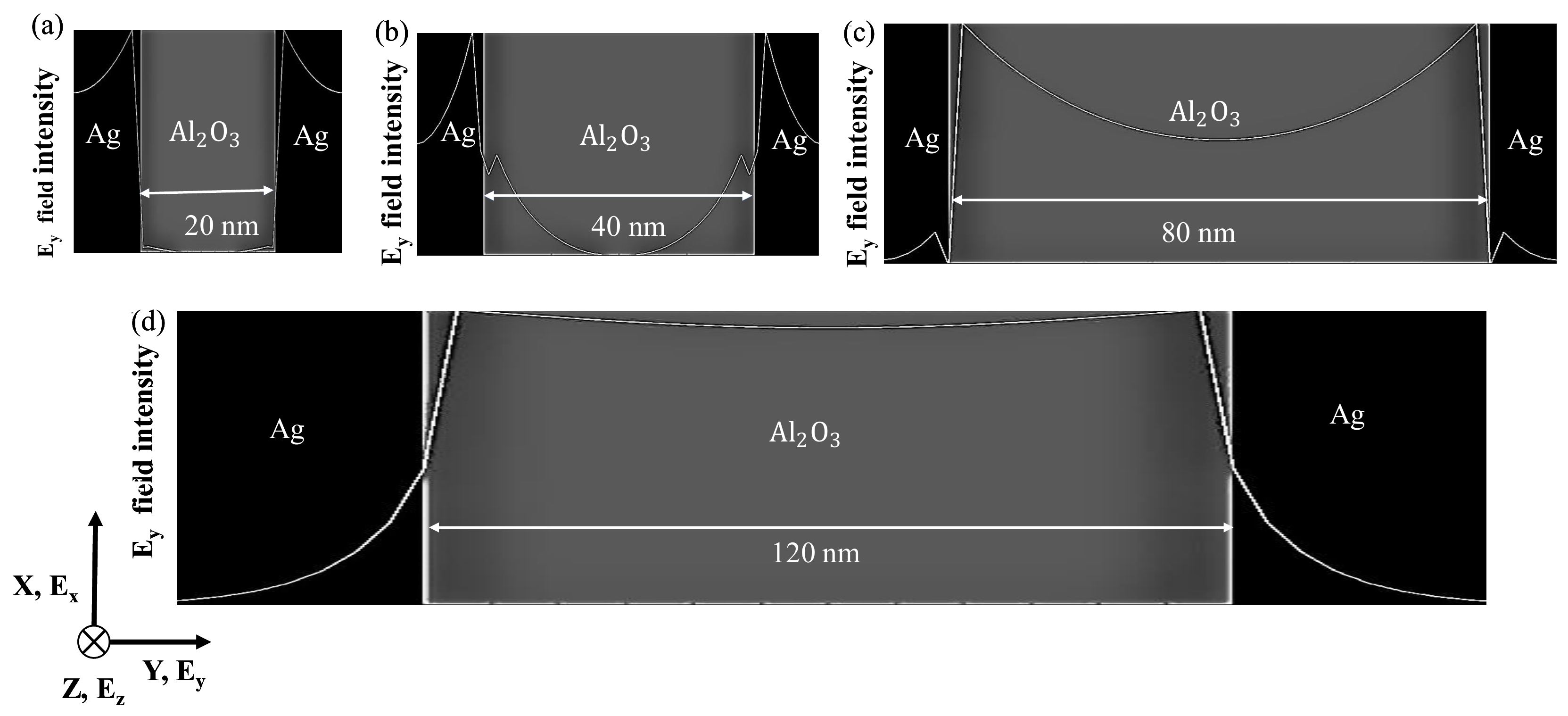}
	\caption{Transverse E$_{y}$ field profile of the MDM waveguide for various dielectric width (w) and dipole emitter wavelength ($\lambda$). (a) w = 20 nm \& $\lambda$ = 331 nm, (b) w = 40 nm \& $\lambda$ = 352 nm, (c) w = 80 nm \& $\lambda$ = 437 nm and (d) w = 120 nm \& $\lambda$ = 521 nm.}	
	\label{fig4}
\end{figure*}

where K$_{o}$ is the propagation vector in free space, w is the width of the dielectric in MDM waveguide, $\lambda$ is the wavelength of the EM wave, $\epsilon_{d}$, $\epsilon_{m}$ are the permitivitty of dielectric and metal respectively. The component of wave vector perpendicular to the interface which quantifies the confinement of the wave can be calculated using $K_{spp}$ as following \cite{Maier2007}.
\begin{equation}
K_{y} = \sqrt{K_{spp}^{2}-\epsilon_{d}K_{o}^{2}}
\label{eq7}
\end{equation} 
The propagation length (L$_{spp}$) is defined as the distance by which the SPP field intensity drops to 1/e, which is theoretically \cite{Maier2007} given by $ (2Im[K_{spp}])^{-1 } $. The component of wave vector perpendicular to the interface defines another quantity called evanescent decay length \cite{Maier2007}, which is equal to 1/K$_{y}$ and quantifies the confinement of the wave. 

\begin{table}\small
	\caption { Calculated values of  key parameters for the analysis of SPP in the MDM waveguide such as  effective mode length (L$_{eff}$), quality factor (Q), modal volume (V), group refractive index (n$_{g}$), plasmonic Purcell factor (F$_{p}$) and the coupling efficiency ($\beta$).}	
		\begin{threeparttable}
			\begin{tabular}{ lccccccccccc }
				\hline
				w (nm)\tnote{a} & $\lambda_{max}$ (nm)\tnote{b} & $\delta$ (nm)\tnote{c} & L$_{spp}$ ($\mu$m)\tnote{d} & L$_{eff}$ (nm)\tnote{e} & Q\tnote{f} & V  (nm$^{3}$)\tnote{g} & $\frac{V}{(\lambda/n)^{3}}$\tnote{h} & n$_{g}$\tnote{i} & F$_{p}$\tnote{j} & $\beta$\tnote{k} \\ \hline
				1              & 307$^{*}$                     & 1.726                  & 0.1                 & 0.863                  & 58.32      & 149                   & 0.0001                                & 28.55                                & 31974           & 99.99\%          \\ 
				5              & 310$^{*}$                     & 3.91                   & 0.231               & 1.958                  & 60.83      & 1777                  & 0.002                                 & 12.99                               & 2879           & 99.96\%          \\ 
				10             & 313$^{*}$                     & 5.64                   & 0.342               & 2.823                  & 64.03      & 5460                  & 0.005                                 & 9.34                                & 1022            & 99.90\%          \\ 
				20             & 331                           & 8.65                   & 0.547               & 4.326                  & 70.52      & 20498                 & 0.015                                 & 4.81                                & 249            & 99.60\%          \\ 
				30             & 331                           & 10.59                  & 0.70                & 5.298                      & 77.40      & 39445                 & 0.029                                 & 3.75                                & 129            & 99.23\%          \\ 
				40             & 352                           & 13.40                  & 0.92                & 6.7011                  & 84.54      & 82693                 & 0.051                                 & 3.04                                & 74.24            & 98.67\%          \\ 
				50             & 352                           & 14.98                  & 1.072               & 7.492                & 91.71      & 120372                & 0.074                                 & 2.61                                & 51.00            & 98.07\%          \\ 
				60             & 376                           & 18.06                  & 1.331               & 9.032                & 99.37      & 217305                & 0.1103                                & 2.24                                & 34.43            & 97.17\%          \\ 
				70             & 404                           & 21.55                  & 1.629               & 10.779               & 107.11     & 378626                & 0.15504                               & 1.97                                & 24.51            & 96.08\%          \\ 
				80             & 437                           & 25.62                  & 1.974               & 12.814               & 114.74     & 648408                & 0.20978                               & 1.76                                & 18.11            & 94.76\%          \\ 
				90             & 437                           & 27.18                  & 2.164               & 13.592                & 122.62     & 799868                & 0.25878                               & 1.60                                & 14.68            & 93.62\%          \\ 
				100            & 475                           & 31.97                  & 2.576               & 15.989               & 130.04     & 1317242               & 0.33189                               & 1.46                                & 11.45            & 91.97\%          \\ 
				110            & 521                           & 37.97                  & 3.051               & 18.878               & 136.672    & 2175310               & 0.41531                               & 1.35                                & 9.15             & 90.14\%          \\
				120            & 521                           & 39.43                  & 3.280               & 19.718               & 144.754    & 2551046               & 0.48704                               & 1.26                                & 7.80            & 88.64\%          \\
				130            & 576                           & 46.47                  & 3.840               & 23.238                & 150.203    & 4148286               & 0.58609                               & 1.19                                & 6.48            & 86.64\%          \\ 
				140            & 576                           & 48.23                  & 4.091               & 24.115                 & 158.26     & 4758647               & 0.67233                               & 1.12                                & 5.65            & 84.97\%          \\ 
				150            & 644                           & 57.09                  & 4.750               & 28.546               & 161.75     & 7741885               & 0.78262                               & 1.07                                & 4.85           & 82.92\%          \\ \hline
			\end{tabular}
			\label{table1}
			\begin{tablenotes}
				\tiny
				\item[a]w is the   dielectric-width of the MDM waveguide in nm,
				\item[b]$\lambda_{max}$ is the wavelength having maximum transmission for a fixed width of the MDM waveguide,
				\item[c]$\delta$ is the theoretical decay length in nm,
				\item[d] L$_{spp}$ is the theoretical propagation length in $\mu$m, [e]L$_{eff}$ is the effective mode length ($\delta$/2),
				\item[f]Q is the quality factor,
				\item[g]V is the modal volume of MDM waveguide,
				\item[h] ${V}/{(\lambda/n)^{3}}$ is the ratio of the modal volume in MDM waveguide to dielectric waveguide with n equal to refractive index of Al$_{2}$O$_{3}$,
				\item[i]n$_{g}$ is the group refractive index (n$_{g}$ $\approx$ |n$_{eff}$| ),
				\item[j] F$_{p}$ is the plasmonic Purcell factor and
				\item[k] $\beta$ is the coupling efficiency. 
				\item The asterisk(*) values of $\lambda_{max}$ have been computed by extrapolating the   dielectric-width vs transmission $\lambda_{max}$ plot.
			\end{tablenotes}
		\end{threeparttable}

\end{table}
Fig.~\ref{fig2} (b) \& (d) shows the simulation results for the transmission spectrum for various dielectric-width (w) ranging from 150 nm to 20 nm in steps of 10 nm. From the transmission vs wavelength plot, we find the wavelength for maximum transmission ($\lambda$$_{max}$). In Fig.~\ref{fig2} (d) we observe a shift in $\lambda$$_{max}$ towards the smaller wavelength (blue shift) as the   dielectric-width (w) is reduced continuously from 150 nm to 20 nm.  To have a good transmission rate for the MDM wave-guide we need a large value of propagation length and since propagation length decreases with reducing   dielectric-width; we found that transmission too decreases with reducing   dielectric-width.  Once $\lambda$$_{max}$ is known, we calculate the theoretical propagation length (L$_{spp}$) which is equal to 1/2Im[K$_{spp}$], from equation~\ref{eq6} and surface plasmon polariton decay length ($\delta$) from the real part of K$_{spp}$ using equation~\ref{eq7}. The calculated values of L$_{spp}$ and $\delta$ are plotted in Fig.~\ref{fig2} (c). From the FDTD simulation, we find that propagation length (L$_{spp}$) decreases with reducing w and for w =  140 nm the maximum propagation length (L$_{spp}$) of 3.98 $\mu$m is found. The theoretical calculation for L$_{spp}$ is very close to the simulation results and the decay length ($\delta$) gets as small as 9 nm for w = 20 nm. A similar relation was also shown in references, \cite{Dionne2005} and \cite{Noghani2014}.

The effect of changing the dielectric-width (w) on the confinement and the propagation length for our Ag- Al$_{2}$O$_{3}$-Ag MDM waveguide, has been studied using equation~\ref{eq6} \cite{Collin2007}. From equation~\ref{eq6} and from Fig.~\ref{fig3}, we observe that the Re[K$_{spp}$] is inversely proportional to the dielectric-width (w), so the smaller the dielectric-width larger is the value of K. But we are not interested directly in the value of K$_{spp}$, what we want is a long propagation length and small decay length. Fig.~\ref{fig3} also plots the propagation length (L$_{spp}$) and decay length ($\delta$) for various dielectric-width and tells us that we can get good confinement by using smaller dielectric-width but at the same time will lose long propagation length. Hence it means that we cannot have the best of both the properties simultaneously.Although, confinement of light beyond diffraction limit is advantageous for the field of plasmonics  however a trade-off due to  loss  as a result of  the confinement needs to be taken into account. 

It is known that surface plasmon polaritons can only originate for the Transverse Magnetic (TM) mode of electromagnetic waves \cite{Maier2007}. These TM plasmonic modes are characterized by the E$_{y}$ field distribution as either symmetric mode TM$_{1}$ (even modes) or anti-symmetric mode TM$_{0}$ (odd modes). For TM$_{1}$ mode the  E$_{y}$ field is an even function with field maxima at each metal-dielectric interface and for  TM$_{0}$ mode the E$_{y}$ field is an odd function with field maxima at one metal-dielectric interface and field minima at another interface \cite{Dionne2010, Maier2007}. Fig.~\ref{fig4} shows the transverse E$_{y}$ field profile at the two interfaces which are symmetrical and hence the propagation mode is  TM$_{1}$. A similar pattern for the symmetric mode was also observed in Ref~\cite{Dionne2010}.

Using the calculated values for L$_{spp}$ and $\delta$, we derive the other key parameters for the analysis of SPP in the waveguide. These parameters are effective mode length (L$_{eff}$), quality factor (Q), modal volume (V), group refractive index (n$_{g}$), plasmonic Purcell factor (F$_\text{p}$) and the coupling efficiency ($\beta$) as shown in Table.~\ref{table1}. The approach taken to calculate these parameters is discussed in Appendix D.
The coupling efficiency ranges between 82 \% to 99 \% which is very efficient for the integration of SPS to nano waveguides. Coupling efficiency of quantum emitters and SPP strongly depends on dipole orientation. Maximum coupling was observed earlier \cite{BermudezUrena2015}  when dipole orientation is perpendicular the wave-guide axis (as is the case for our calculations i.e. in XY plane and along Y axis as shown in \ref{fig:1} a-b). Furthermore, the Bermudez et al \cite{BermudezUrena2015} had shown that the coupling was shown to be zero when dipole was oriented along X and Z axis. The SPSs like WSe$_{2}$, hBN (hexagonal Boron Nitride), MoS$_{2}$ monolayer and NV (Nitrogen Vacancy) center in nanodiamond crystals can be integrated deterministically by  all-dry viscoelastic stamping \cite{CastellanosGomez2014}. Deterministic integration of SPS to plasmonic waveguide has been shown by several researchers \cite{Siampour2017a,BermudezUrena2015,Blauth2017,Siampour2017,Siampour2018,Cai2017,Dutta2018}. Along with the coupling of SPS to a plasmonic waveguide, minimum loss and good confinement are also required for integrated photonic devices. For dielectric waveguides, the mode volume is the order of $(\frac{\lambda_{0}}{n})^{3}$ where n is the refractive index of the medium. But for our plasmonic waveguide, the modal volume ranges between 0.01 \% to 78 \% of  $(\frac{\lambda_{0}}{n})^{3}$ which is very smaller than that for the dielectric waveguide \& hence proves the subwavelength confinement of the mode. The plasmonic Purcell factor is found to be inversely proportional to modal volume (V) and ranging between 4 to 31974 for various dielectric-width as shown in Table~\ref{table1}.

\begin{table}
	\caption { Literature survey of Indistinguishability (I), efficiency ($\eta$) and the dephasing rate $ \gamma^{*}$  for various categories of SPS}
		\begin{threeparttable}
			\begin{longtable}{lccc}	
				\hline			
				Category of SPS                                                              &      I      & $\eta$ &   $ \gamma^{*}$    \\ 	\hline
				Self assembled InAs/GaAs quantum dots coupled to a \\ photonic crystal cavity \cite{Grange2015}                       &    72\%     & 8.8 \% &  116.6$ \gamma $   \\
				Single silicon vacancy (SiV) center in a nano-diamond \\coupled to a fiber cavity \cite{Grange2015}                     &    81\%     & 3.5\%  & 3437.5  $ \gamma $ \\
				Colloidal quantum dots in coupled cavities\\ (optimal) \cite{Saxena2019}                                  &    90\%     &  0.24\%  &    83000$\gamma$    \\
				Colloidal quantum dots in coupled cavities\\ (experimental) \cite{Saxena2019}                                &    63\%     &  0.15\%  &   83000$\gamma$    \\
				GaAs QD \cite{Reindl2019,Nawrath2019}                                                   &    90\%     &   -    &         -          \\
				GaAs with new broadband photonic structure,\\ CBR-HBR\tnote{a}~  \cite{Liu2019}                      &    90\%     &  85\%  &      -          \\
				 InGaAs QD coupled  to elliptical micro pillars\\ Bragg Grating devices \cite{Wang2019a}                      &    97\%     &  60\%  &         -          \\
				Si vacancy center in diamond at room temperature\\ with cascaded cavity systems  \cite{Choi2019}                      &   31.5\%    &  98.7\% &  10$^{4} \gamma$   \\
				Quantum dots coupled to single/multimode -ridged waveguide \cite{Dusanowski2019}                             & 95\%/97.5\% &   -    &         -          \\
				BBO(Barium borate)-SPDC source \cite{Kaltenbaek2006}                                           &    83\%     &   -    &         -          \\
				Si vacancy center in diamond, (5 K) \cite{Sipahigil2014}                                         &    72\%     &   -    &         -          \\
				Atom-cavity system (Rb atom - cavity) \cite{Legero2004}                                          &    90\%     &   -    &         -          \\
				Dibenzanthanthrene (DBATT) molecule  incorporated in Shpolskii matrices\\ of n-tetradecane  \cite{Ahtee2009}                &      -      &  30\%  &         -          \\
				WSe$_{2}$monolayer onto a SiN waveguide \cite{Peyskens2019}                                        &     7\%     &  93\%  &         -          \\
				Potassium
				titanyl phosphate (aKTP)  crystal- parametric\\ downconversion (PDC) source \cite{Graffitti2018} &    98\%     &   -    &         -          \\
				Quantum dots coupled to a cavity ( 4 K) \cite{Iles-Smith2017}                                       &    99\%     &  96\%  &         -\\ \hline
			\end{longtable}
			\begin{tablenotes}	
				\footnotesize
				\item [a] CBR-HBR : Circular Bragg resonators on highly efficient broadband reflectors 		
			\end{tablenotes}
			\label{table2}
		\end{threeparttable}
\end{table}

\section{Calculation for Indistinguishability \& Extraction effficieny}

One can achieve good extraction efficiency and indistinguishability by coupling the SPS with a nanoplasmonic cavity which is evanescently coupled to the SPP plasmonic mode of the waveguide. Spontaneous emission by vacuum fluctuations for a two-level quantum system exhibits pure indistinguishable photons. Often the dephasing takes place which relates to the quantum decoherence, and in that case the indistinguishability is represented by the following equation
\begin{equation}
I = \frac{\gamma}{\gamma+\gamma^{*}}
\label{15}
\end{equation}
where $\gamma$ represents the quantum emitter decay rate and the $\gamma^{*}$ the dephasing rate \cite{Grange2015}. The value for $\gamma^{*}$ is around 10$^{4}$ $\gamma$, hence it can be seen from the formula that indistinguishability is very low, of the order of 10$^{-4}$, for an emitter  in free space or not coupled to a cavity. \cite{Saxena2019,Choi2019}. However,  coupling SPS inside an optical cavity results in an increase of both the extraction efficiency and the indistinguishability \cite{Gazzano2013, Gerard1998, Albrecht2013, Choi2019}, the increment in I is due to the modification of the local density of the electromagnetic states (LDOS). In the cavity quantum electrodynamics (cavity-QED) picture, the relevant parameters for the drastic change in the indistinguishability are the emitter-cavity coupling rate (g), $\kappa$ the cavity decay rate and Q the quality factor \cite{Saxena2019, Grange2015, Choi2019}. The approach taken to calculate these parameters is discussed in Appendix E.

\begin{table}
	\caption{Values of various key parameters for the analysis of SPS and nanoplasmonic cavity coupled system.}
	\label{table:3}
\centering
\small
		\begin{threeparttable}			
			\begin{tabular}{ ccccccccccccc }
				\hline
				SPS\tnote{a}                               & $\lambda$ (nm)\tnote{b} & Q\tnote{c} &    p$_{d}$(D)\tnote{d}    & g (THz)\tnote{e} & $\kappa$ (THz)\tnote{f} & $\gamma$ (MHz)\tnote{g} & $\gamma$$^{*}$ (MHz)\tnote{h} & R (THz)\tnote{i} & I\tnote{j} & $\eta$\tnote{k} & (I$ \times  $$\eta$)\tnote{l} &  \\ \hline
				Methylene blue$^{*}$\cite{Chikkaraddy2016} &           665           &    15.9    & 3.8\cite{Chikkaraddy2016} &      82.44       &           178           &          15.34          &  $\gamma$ $\times$ 10$^{4}$   &      152.44      &   0.9987   &     0.9998      &            0.9985             &  \\
				CsPbI$_{3}$$^{*}$\cite{Park2015c}          &           660           &     8      &     3.5\cite{Sun2018}     &      71.26       &           356           &          13.31          &  $\gamma$ $\times$ 10$^{4}$   &      56.904      &   0.9907   &     0.9999      &            0.9906             &  \\
				CdSe\cite{Aichele2004}                     &           510           &     8      & 10\cite{Nakabayashi2014}  &       230        &           461           &         235.61          &  $\gamma$ $\times$ 10$^{2}$   &       458        &   0.9999   &     0.9999      &            0.9998             &  \\
				InGaAlAs\cite{Rakhlin2018}                 &           680           &     8      &   29\cite{Eliseev2000}    &       582        &           346           &         835.96          &  $\gamma$ $\times$ 10$^{2}$   &       3911       &   0.9996   &     0.9999      &            0.9995             &  \\
				WSe$_{2}$\cite{Koperski2015}               &           713           &     8      &     16\cite{Jin2016}      &       518        &           330           &         220.74          &  $\gamma$ $\times$ 10$^{2}$   &       3249       &   0.9999   &     0.9999      &            0.9998             &  \\
				InGaN\cite{Deshpande2013}                  &           450           &     8      &  21\cite{Ostapenko2010}   &       312        &           523           &         1512.58         &  $\gamma$ $\times$ 10$^{2}$   &       743        &   0.9996   &     0.9999      &            0.9995             &  \\
				hBN \cite{Xia2019}                         &           594           &     8      &    0.65 \cite{Xia2019}    &      13.94       &           396           &         0.6291          &  $\gamma$ $\times$ 10$^{4}$   &      1.961       &   0.9873   &     0.9979      &            0.9852             &  \\
				NV in diamond \cite{Tamarat2006}           &           637           &     8      &  1.3 \cite{Tamarat2006}   &      26.94       &           369           &          2.03           &  $\gamma$ $\times$ 10$^{4}$   &       7.86       &   0.9898   &     0.9995      &            0.9893             & \\ \hline
			\end{tabular}
			\label{table3}
			\begin{tablenotes}
				\footnotesize
				\item[a] SPS is the Single Photon Source,
				\item[b] $\lambda$ is the wavelength of the mentioned SPS,
				\item[c] Q is the quality factor,
				\item[d] p$_{d}$ is the dipole moment strength for the SPS shown in D (Debye) which is equal to 3.3 x 10$^{-30}$ C.m,
				\item[e] g is the emitter cavity coupling rate,
				\item[f] $\kappa$ is the cavity decay rate,
				\item[g] $\gamma$ is the emitter decay rate,
				\item[h] $\gamma$$^{*}$ is the decoherence rate,
				\item[i] R is the population transfer rate between emitter and cavity,
				\item[j] I is the indistinguishability,
				\item[k] $\eta$ is the extraction efficiency and
				\item[l] I $\times$ $\eta$ is the product of indistinguishability (I) and efficiency $\eta$ , which should be 1 for an ideal SPS. 
			\end{tablenotes}
		\end{threeparttable}

\end{table}

Many researchers have probed the use of nanoplasmonic cavities and antennas to enhance the light-matter interactions using structures such as bow-tie antenna and NPoM (nanoparticle on a mirror) coupled with SPSs \cite{Kongsuwan2017, Chang2007, Kleemann2017, Fang2011, Conteduca2017, Lu2013, Chikkaraddy2016}. We have calculated the indistinguishability and extraction efficiency for SPSs coupled with an NPoM plasmonic cavity as fabricated in \cite{Chikkaraddy2016} which in turn is coupled evanescently with our MDM waveguide.  The NPoM cavity consists of a spherical  gold  nanoparticle of 40 nm diameter on the top of a 70 nm thick gold film separated by a 0.9 nm molecular spacer \cite{Chikkaraddy2016, Aravind1982}.  We have assumed that the cavity and emitter are resonant with each other making $\omega_{c}$ = $\omega$$_{e}$. As the plasmonic cavity, we have selected the NPoM because it is possible to change the resonant wavelength of the cavity by changing the shape and size of the nanoparticle in the plasmonic cavity \cite{Chikkaraddy2016}.
 
The SPSs we have used for the calculations are methylene blue single molecule \cite{Chikkaraddy2016}, CsPbI$_{3}$\cite{Park2015c}, hexagonal Boron Nitride (hBN) \cite{Xia2019}, CdSe \cite{Aichele2004}, InGaAlAs \cite{Rakhlin2018}, InGaN \cite{Deshpande2013}, WSe$_{2}$ \cite{Koperski2015} and Nitrogen vacany (NV) in diamonds \cite{Tamarat2006} ; out of these Methylene blue, CsPbI$_{3}$, NV in diamond \& hBN are room-temperature single-photon sources ($\gamma^{*}$ of the order of 10$^{4}$ times $\gamma$) while the rest require a cryogenic set up to operate as SPS($\gamma^{*}$ of the order of 10$^{2}$ times $\gamma$). 
For the calculation of indistinguishability (I), we have kept the quality factor (Q) equal to 8 and modal volume (V) equal to 40 nm$^{3}$ for all the SPS, apart from the methylene blue for which the values quoted in Chikkaraddy et.al.\cite{Chikkaraddy2016} have been used which are Q = 15.9 and V =  35 nm$^{3}$. For the NPoM cavity we found that the Purcell factor is of the order of 10$^{6}$. 
For all these SPSs the emitter wavelength and the cavity decay rate ($\kappa$) is shown in Table.~\ref{table3}.
The emitter cavity coupling strength  "g" is calculated using the dipole strength (p$_\text{d}$) as quoted in Table.~\ref{table3}, keeping cos$\theta_{d}$ as 1. From the calculated values of $\kappa$, $\gamma$ and $\gamma^{*}$ and g, we found that quantum emitters such as  methylene blue single molecule,  CdSe,  InGaAlAs,  WSe$_{2}$, and  InGaAlAs are in the strong coupling regime and  CsPbI$_{3}$, hexagonal Boron Nitride (hBN) and Nitrogen Vacancy (NV) in diamond  are in the weak coupling regime. Our calculated value of 2g for methylene blue single molecule matches with the experimental value of rabi splitting as quoted in \cite{Chikkaraddy2016}.  The valuee for $\gamma$,  the emitter decay rate was calculated using the transition rate formula given by Fermi's golden rule \cite{Fox2006} \& $\gamma$$^{*}$ the dephasing rate is taken as 100 times $\gamma$ for low temperature  (4K) \cite{Peyskens2019}  and 10000 times $\gamma$ for room temperature \cite{Grange2015}. We have calculated the values for I \& $\eta$ as shown in Table.~\ref{table3} for both strong and weak coupling using the equations for indistinguishability and extraction efficiency as discussed in Appendix E. For both the case of strong and weak coupling  in cavity-emitter system, we find that I and $\eta$ are  close to 99 \%.  These values are adequately high as compared to the emitters reported earlier in the Table~\ref{table2}.  We believe this makes  these emitters better SPS to be used for quantum  technology application. The increase in the single photon indistinguishability and extraction efficiency for a weak coupling between the cavity and emitter is due to the high Purcell factor for the plasmonic cavities. Switching from  weak coupling regime to the strong coupling results in an increase in the Purcell factor (F$_{p}$) as F$_{p}$ is directly proportional to g$^{2}$. However for strong coupling  there is an upper bound on the effect of Purcell factor on the indistinguishability and extraction efficieny due to the clamping of the effective decay rate of the quantum dot\cite{Kaer2013}. The use of strong coupling regime between emitter and cavity to enhance the single photon properties has been experimentally reported in \cite{Gerhardt2019, Hennessy2007,Press2007}. 

\section{Quantum information processing}

\begin{figure}
	\centering
	\includegraphics[width=0.6\linewidth]{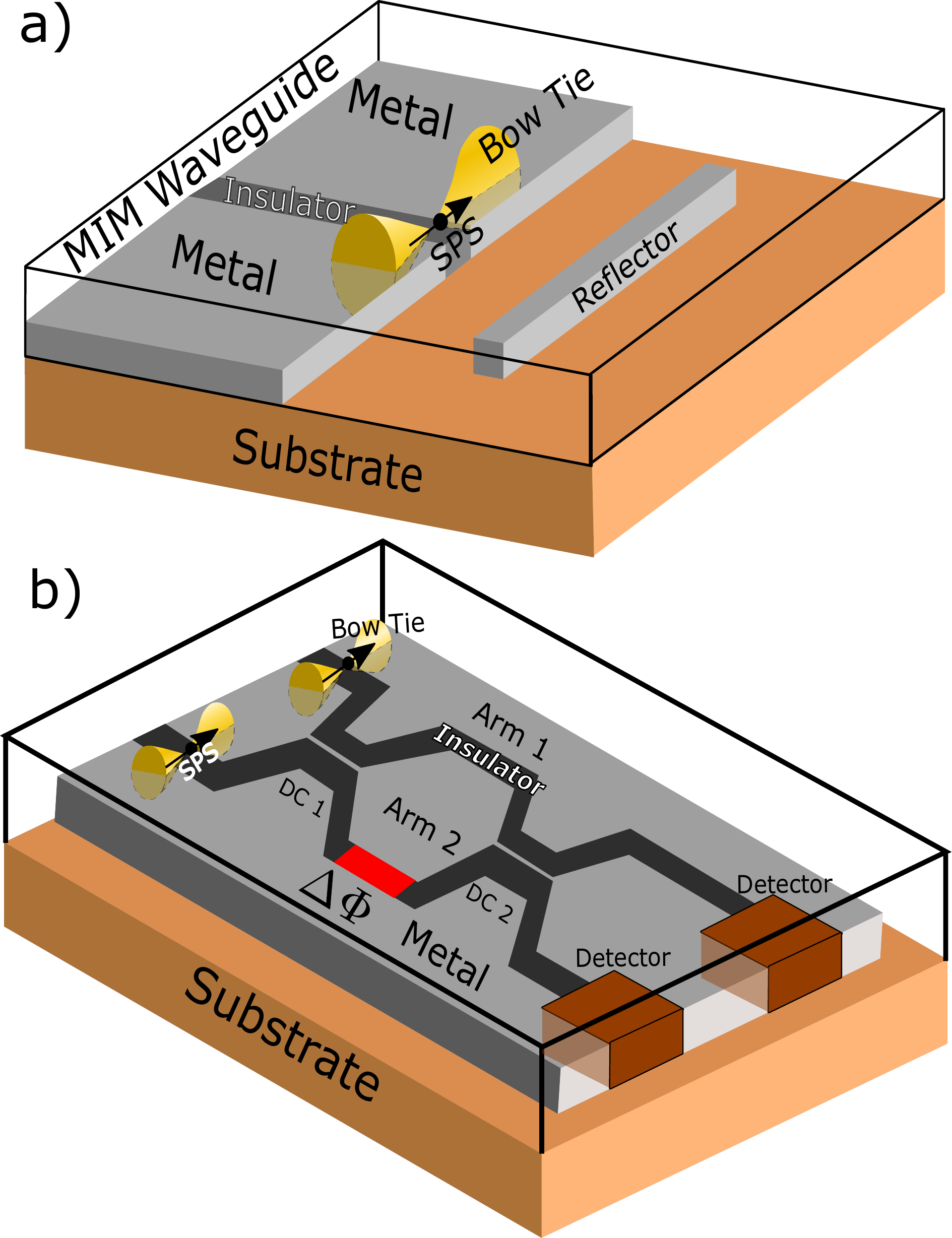}
	\caption{a)  Proposed schematics  for a quantum logic circuit: Deterministically placed hybrid bow-tie nanoantenna single photon sources (SPS) in a metal-dielectric-metal(MIM) waveguide. b) Mach-Zehnder interferometer acting as  controlled quantum logic gate, deterministically placed antenna-SPS metal-dielectric-metal waveguide sends polarized qubits, directional coulers (DC1 and DC2) acting as the beam splitter process the information and   qubit  is registered by single photon detector (superconducting nanowires etc). The phase shift $\Delta$ $\Phi$ is introduced by a non-linear material placed inside the wave guide ( red region).}
	\label{fig:mz}
\end{figure}

Mach-Zehnder interferometer circuits using MDM waveguide and single photons can lead to quantum computers \cite{MichaelA.Nielsen2010}. Using nonlinear Kerr medium ( Fig.~\ref{fig:mz}, red region) such as MEH-PPV [poly(2-methoxy-5-(28-ethylhexyloxy)-PPV)] having $ n_{2} $ $ 1.8 \times 10^{-13} cm^2/W$ at 1064 nm   wavelength \cite{Bader2002, Pu2010}  and InAs quantum dots \cite{Nakamura2004}  phase shift $\Delta$ $\Phi$ pf  ($ \pi $) can be achieved. On-resonant quantum dots and cavity in one arm can also lead to the phase shift of $\pi$ \cite{Androvitsaneas2019}. For cross-phase modulation (XPM), the relative phase shift of $ \pi  $  in both arms require firstly,  one of the arms to be longer. Secondly, the nonlinear materials hold have larger nonlinear susceptibilities $\chi^3$, and thirdly, the sources of the single photons should be brighter to enhance the intensity (I) dependent refractive index ($\Delta$  n = $ n_{2}$ I), $ n_{2}$ is a nonlinear coefficient. Our metal-dielectric-metal waveguide serves longer surface plasmon polariton propagation length helping solve the first problem. Choosing higher non-linear coefficient (n$_{2}$) materials will solve the second problem.  While Purcell enhanced single-photon source, for example, bow tie or NPoM coupled SPS is a potential candidate for brighter and faster single-photon source solving the third problem. Our proposed scheme to build quantum logic circuits using a nanoantenna enhanced SPS is shown in Fig.~\ref{fig:mz}. 

\section{Conclusion}
In conclusion, the operation of plasmonic waveguide based on SPP in photonic integrated
circuits has been studied. The key parameters that are needed for full-scale implementation of
plasmonic waveguides such as propagation length, decay length, coupling efficiency (between the	waveguide and quantum emitter), and plasmonic Purcell factor for MDM waveguide have been analyzed using
FDTD simulations and analytical methods. We found the coupling efficiency to be greater than 82\% for dielectric-width (w) in the range of 20 nm - 150 nm and the plasmonic Purcell factor increasing with decreasing w, reaching as high as 31974 for w = 1 nm from 4.85 for w = 150 nm. We found the maximum propagation length (L$_{spp}$) of 3.98 $\mu$m from the simulations for dielectric-width equal to 140 nm at $\lambda$ equal to 576 nm. Along with this, we found that the decay length ($\delta$) is directly proportional to the dielectric-width of the MDM waveguide and ranges between 8 nm to 60 nm for w between 20 nm - 150 nm respectively. Hence it is clear that there exists a trade-off between localization and loss in plasmonic waveguides; the better the confinement (smaller decay length), the lower is the	propagation length.  We also studied the effect on the indistinguishability (I) and extraction efficiency ($\eta$) for several single-photon sources coupled with the NPoM plasmonic cavity. And found indistinguishability close to 99 \% at room temperature and extraction efficiency ($\eta$) around 99 \%. We also proposed the design of a quantum logic gate based on a deterministically placed nano-plasmonic antenna-SPS coupled system with a metal-dielectric-metal waveguide.

\section{Acknowledgment}

We acknowledge Birla Institute of Technology, Mesra, Ranchi, and the Ministry of Human Resource Development, Government of India, for support through TEQIP‐III and Collaborative Research Scheme (CRS): CRS‐ID:1‐5736483014. We are thankful to Subham Adak for useful discussion and Kostav Konar for technical help.

\appendix
	\section{Interpolation plot for finding the transmission $\lambda_{max} $ values for width 1 nm, 5 nm and 10 nm }

\begin{figure} 
	\centering
	\includegraphics[width=0.8\linewidth]{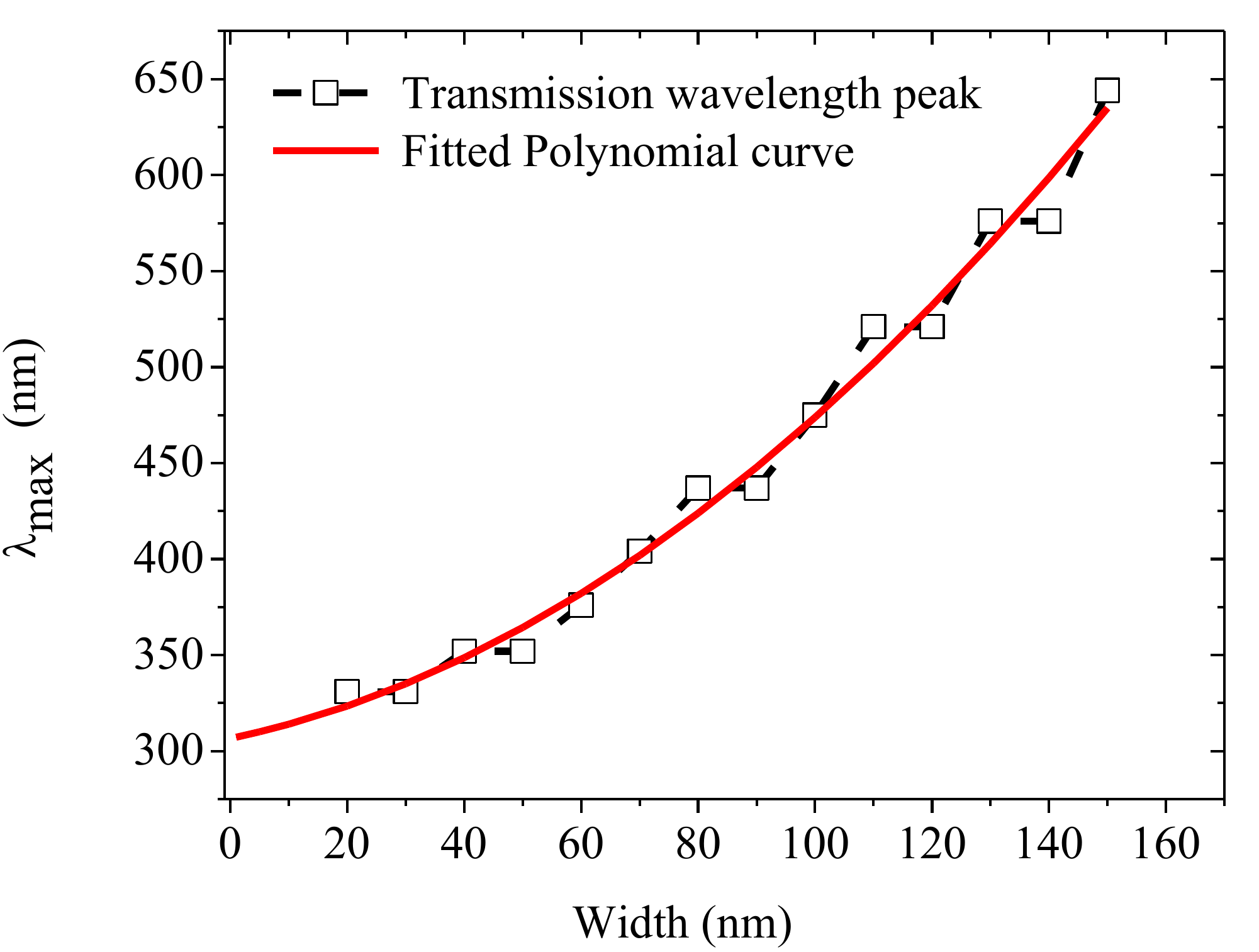}
	\caption{Interpolation plot for width vs Transmission $\lambda_{max}$. }  
	\label{lambdamax}
\end{figure}
In Table.~\ref{table1} of the paper the asterisk($\star$) values of $\lambda$$_{max}$ have been found using the folowing interpolation plot. From the Fig.~\ref{lambdamax} we get 
\begin{equation}
\lambda_{max} = 306.55+(0.63 w) +(0.01033 w^2)
\end{equation}
where w is the width and using w as 1 nm, 5 nm and 10 nm gives $\lambda_{max}$ as 307 nm, 310 nm and 313 nm respectively. 

\section{Condition for surface plasmon polariton propagation}
\begin{figure} 
	\centering
	\includegraphics[width=0.8\linewidth]{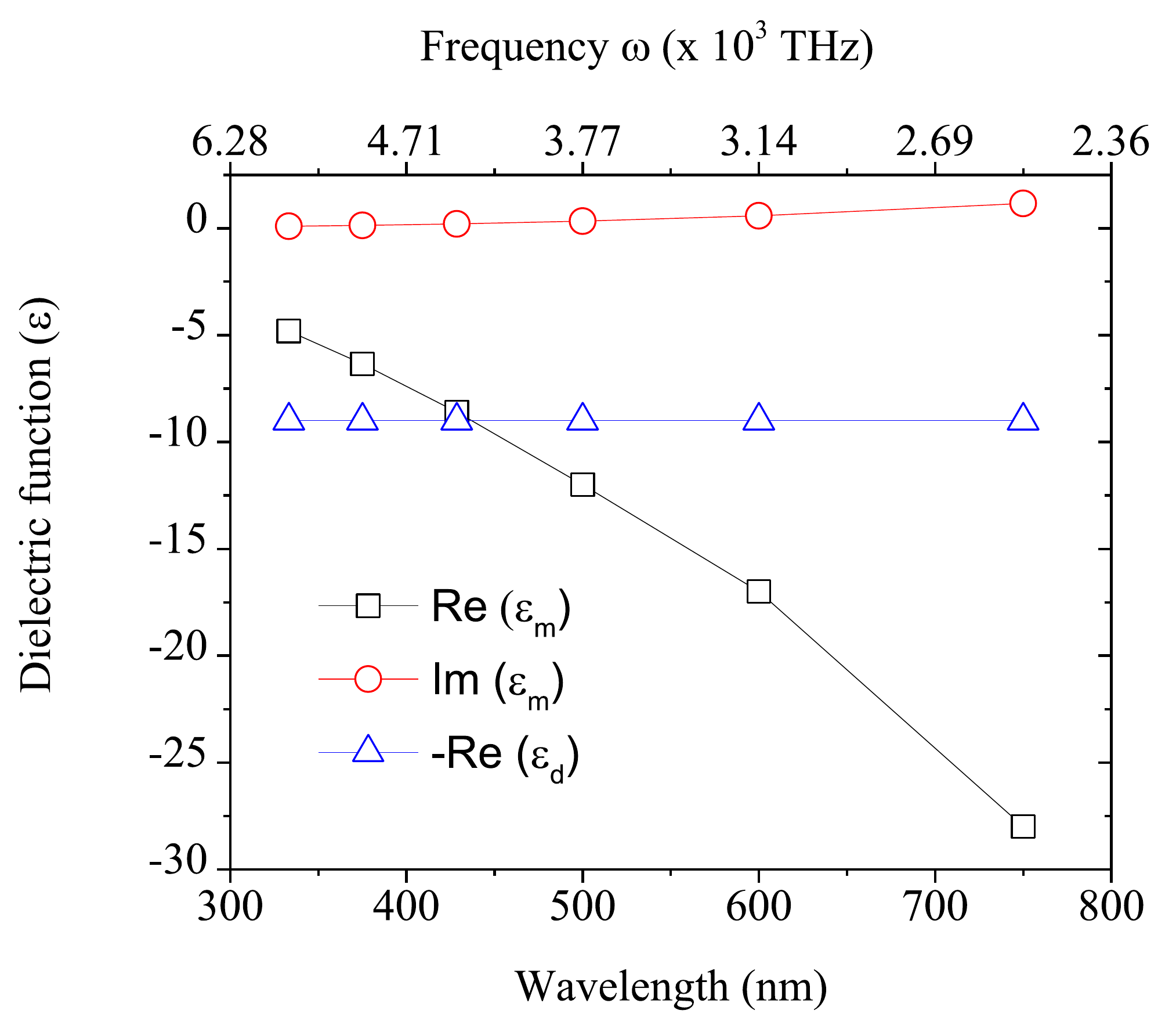}
	\caption{Plot of the complex dielectric function for silver and negative of the  dielectric constant value for Al$_{2}$O$_{3}$.}  
	\label{B.7}
\end{figure}

Typically, used metals for the study of surface plasmon polariton are gold (Au) and silver (Ag). Both these metal are chemically inert, stable, shows excellent tailorable binding to bio-molecules and also have low loss compared to other metals. In this study, we have used Ag which has plasma energy of 9.013 eV ($\omega_{p}$ = $ 1.37  \times 10^4  $ THz) \cite{Herrera2014} and collision frequency $\gamma$ of around 100 THz \cite{Maier2007}.
The dielectric function of Ag as described by Drude model is\cite{Yang2017, Johnson1972}
\begin{equation}
\epsilon(\omega)= 1-\frac{\omega_{p}^{2}}{\omega^{2}+i\gamma\omega}
\end{equation}
The complex dielectric function of Ag has been plotted in Fig.~\ref{B.7}, along with the dielectric constant for  Al$_{2}$O$_{3}$ which is 9.   For $\omega$ $<$ $\omega_{p}$, the real part of the dielectric constant for Ag is negative as it must be for the existence of SPPs and the imaginary part is positive which corresponds to energy loss of propagating SPPs. 
For electromagnetic waves with $\omega$ $>$ $\omega_{p}$, we have a positive $\epsilon$ and hence the electromagnetic waves do not get shielded. So the waves with $\omega$ $>$ $\omega_{p}$ can propagate in the metal. The value of $\omega_{p}$ for metals lies in the ultra-violet range and the dispersion relation for the waves having $\omega$ $>$ $\omega_{p}$ is given by 
\begin{equation}
\omega^{2} = \omega_{p}^{2} + c^{2}K^{2}
\label{25}
\end{equation} 
The plot of $\omega$ and K can be seen in Fig.~\ref{fig2} (a) of the paper.

\section{Simulation results for 100 nm dielectric-width MDM waveguide showing its propagation properties and E field vs Wavelength profile}
\begin{figure} 
	\centering
	\includegraphics[width=0.8\linewidth]{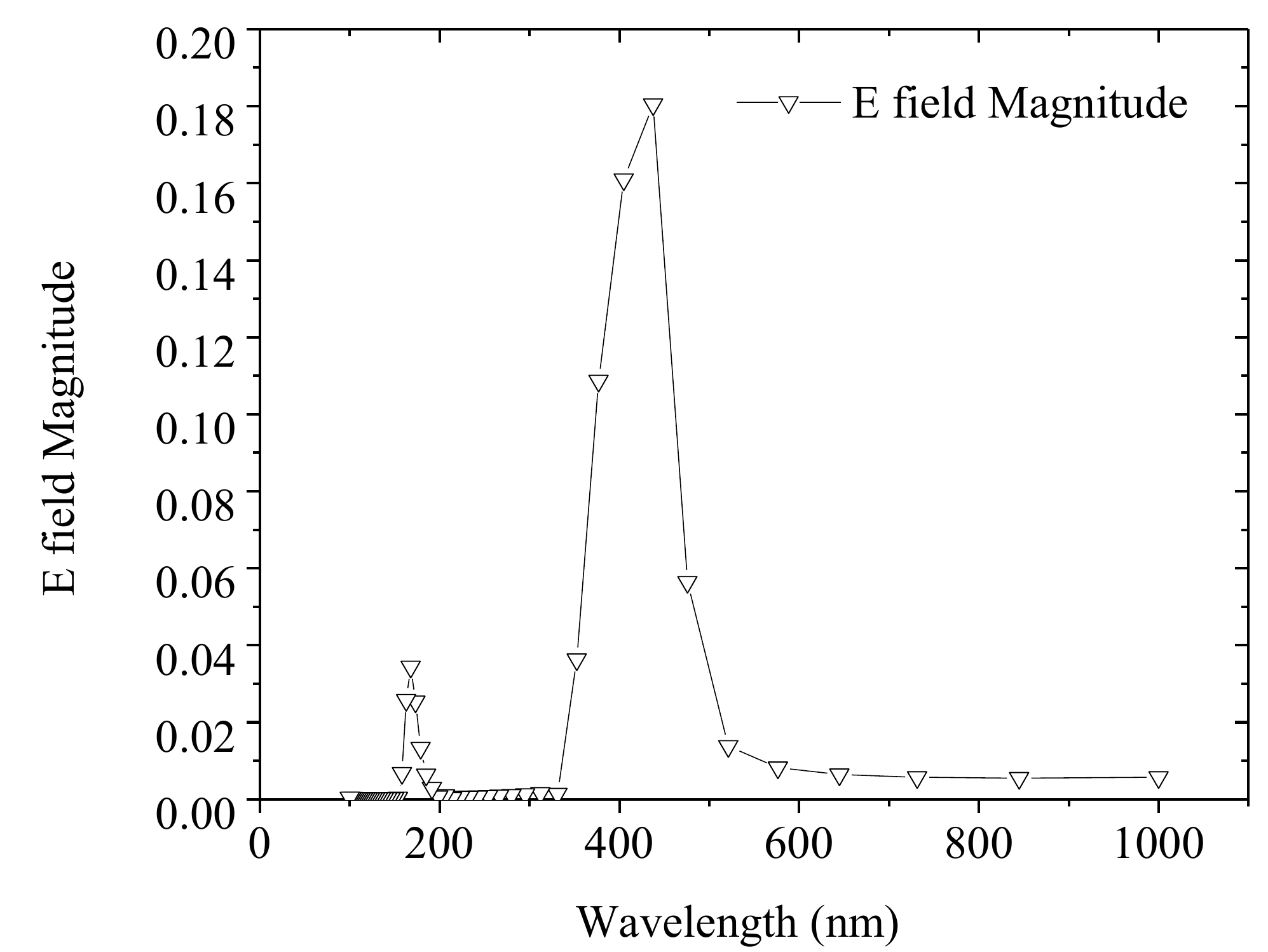}
	\caption{The E field Vs Wavelength profile with SPP peak at $\lambda$ equal to 437 nm, for 100 nm dielectric-width MDM waveguide.}  
	\label{SIFig8}
\end{figure}

\begin{figure} 
	\centering
	\includegraphics[width=0.8\linewidth]{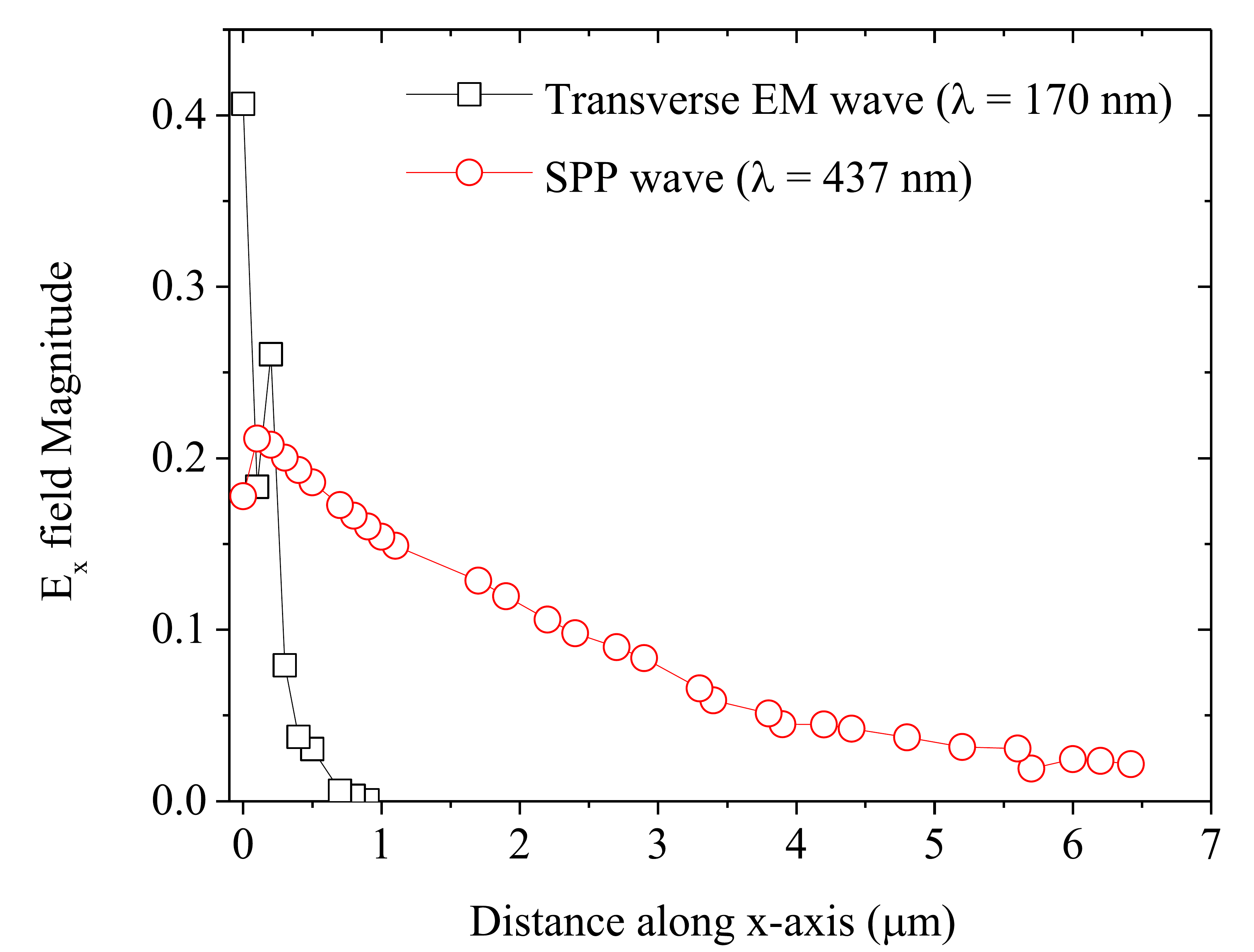}
	\caption{E$_{x}$ field vs distance along x-axis plot for the MDM waveguide showing the propagation property of the SPP for the 100 nm dielectric width MDM waveguide at $\lambda$ equal to 437 nm.}  
	\label{SIFig9}
\end{figure}
For the FDTD simulation the source of the electromagnetic radiation used was a total scattered wave, for which we have selected the range between 100 to 1000 nm. A point size electric field monitor was used to simulate the E field. In Fig.~\ref{SIFig8} the E field vs wavelength profile is plotted; the highest peak is at $\lambda$ equal 437 nm and the lower peak at $\lambda$ equal to 170 nm. The resonance around 437 nm corresponds to the SPP mode of the MDM waveguide. For w = 100 nm, the diffraction limit as calculated by d$\approx$ $\lambda$/2 would give $\lambda$ $\approx$ 200 nm which is very close to the second peak observed at 170 nm which lies in the UV range of the EM spectrum. Along with the field vs wavelength profile, we also simulated the E$_{x}$ field variation in the propagating direction which is the X-axis. In Fig.~\ref{SIFig9} the results of this simulation have been plotted for the two peak wavelength, 170 nm \& 437 nm.  The propagation length (L$_{spp}$) is defined as the distance by which the E-field intensity in the propagation direction drops to 1/e (36.78 \%) of its maximum value. For 100 nm dielectric width of the MDM waveguide at dipole emitter wavelength equal to 437 nm we found that the L$_{spp}$ is equal to 2.6 $\mu$m.	
\begin{figure} 
	\centering
	\includegraphics[width=0.8\linewidth]{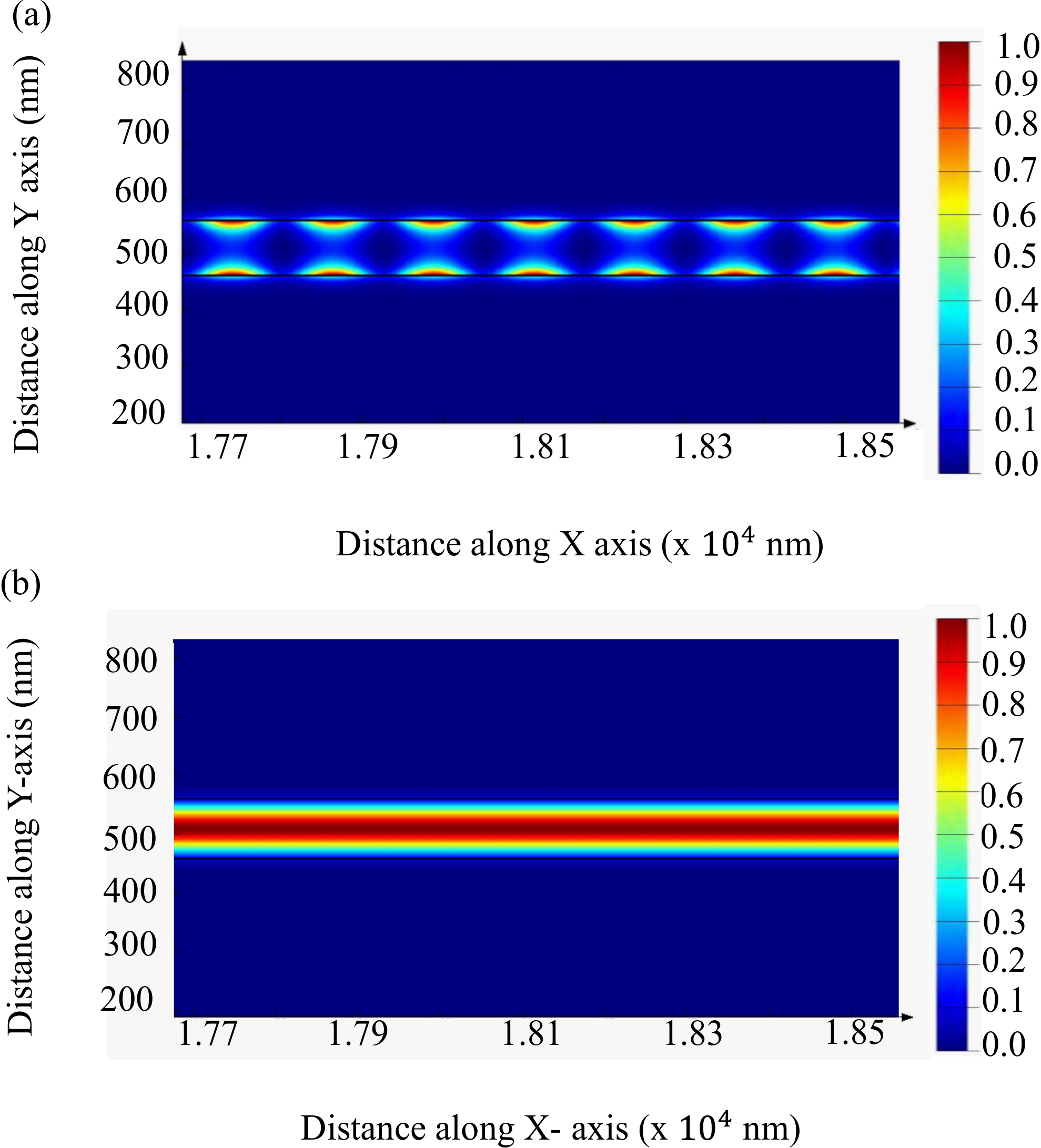}
	\caption{ X-Y field profile of the MDM waveguide for dielectric width (w) equal to 100 nm. (a) Field profile at $\lambda$ = 437 nm \& (b) Field profile at $\lambda$ = 170 nm.
		The color scale bar represents the normalized field intensity.}  
	\label{5}
\end{figure}

The Fig.~\ref{5} shows the simulation results for X-Y field profile; (a) X-Y field profile showing the surface waves (SPP) propagating in the x-direction at $\lambda$ = 437 nm for 100 nm dielectric-width MDM waveguide and (b) X-Y field profile showing the wave propagation at $\lambda$ = 170 nm for 100 nm dielectric-width MDM waveguide. 

\section{Calculation of key parameters for the analysis of SPP in the waveguide}
The environment of a quantum emitter (SPS) alter its spontaneous emission rate given by the Purcell Factor	
\begin{equation}
F_{p} = \frac{3}{4\pi^{2}}(\frac{Q}{V_{c}})(\frac{\lambda}{n})^{3}
\label{eq8}
\end{equation}

Consequently, it's clear that by increasing quality factor (Q) and by lowering the mode volume ($V_{c}$) we can substantially increase the Purcell factor.
The Purcell factor (F$_{p}$) as shown in equation~\ref{eq8} quantifies the emission enhancement of light in an emitter-cavity system. The phenomenon of confinement of light energy in a volume much below the diffraction limit of $(\lambda_{0}/2n)^{3}$ \cite{Maier2007} (where n is the refractive index of the dielectric) in SPP modes also leads to a very small modal volume.
\newline The SPP quality factor Q is defined as the following  \cite{Francs2016}.
\begin{equation}
Q = K_{spp}/ \Delta (K_{spp}) \approx K_{spp}L_{spp}
\label{eq9}
\end{equation}
where K$_{spp}$ represents the SPP wavevector in the propagation direction and $\Delta$ K$_{spp}$ represents the imaginary part of it. The mode volume V for a plasmonic waveguide is \cite{Francs2016}.
\begin{equation}
V = A_{eff} 2L_{spp}
\label{eq10}
\end{equation}
where  L$_{spp}$ is the propagation length and A$_{eff}$ is the effective mode area, which is the measure of the area which a waveguide mode covers in the transverse  direction. For SPP modes the transverse  direction is same as the direction of the confinement of the mode which is represented by the effective mode length \cite{Francs2016}
\begin{equation}
L_{eff}= \frac{\int |(E(z))|^{2} dz}{Max |(E(z))|^{2}} = \int^\infty_0 e^{-2z/\delta}dz = \delta/2
\label{eq11}
\end{equation} 
where $\delta$ is the surface plasmon polariton decay length.
In terms of effective length we can write the effective mode area A$_{eff}$ as (L$_{eff}$)$^{2}$.
The plasmonic Purcell factor for a plasmonic waveguide can thus be defined as \cite{Francs2016}
\begin{equation}
F_\text{p} = \frac{3}{4\pi^{2}}(\frac{\lambda_{em}}{n_{1}})^{3}\frac{Q_{spp}}{V}\frac{\omega_{em}}{V_{g}K_{spp}}
\label{eq12}
\end{equation}

where Q $_{spp}$ is the quality factor, V is the modal volume of the plasmonic waveguide as defined by equation~\ref{eq10}, n$_{1}$ is the refractive index of the dielectric material, V$_{g}$ is the group velocity of SPP.
The group velocity (V$_{g}$) can be calculated by taking the inverse of $\frac{dK_{spp}}{d\omega}$.
And substituiting all the above described factors we get the following formula for the  plasmonic Purcell factor \cite{Francs2016}.
\begin{equation}
F_\text{p} = \frac{3}{4\pi}\frac{(\lambda_{em}/n_{1})^{2}}{A_{eff}}\frac{n_{g}}{n_{1}}
\label{eq13}
\end{equation}
where n$_{g}$ is the group refractive index of the confined mode and n$_{1}$ is the refractive index of the dielectric medium \cite{Francs2016}.
For the practical use of plasmonic waveguide in photonic integrated circuits, the coupling of emitted energy from the dipole or the SPS to the waveguide modes has to be very high. Using the plasmonic Purcell factor we can estimate the coupling efficiency between the dipole emitter and the waveguide modes.  The equation for coupling efficiency ($\beta$) in terms of Purcell factor is given by \cite{Francs2016}.
\begin{equation}
\beta \approx \frac{F_\text{p}}{1+F_\text{p}}
\label{eq14}
\end{equation} 

\section{Cavity QED picture: Calculation of Indistinguishability and Efficiency}
The quantum emitter decay rate $\gamma$ has been calculataed using the transition rate given by the Fermi's golden rule \cite{Fox2006} 
\begin{equation}
\gamma = \frac{p_{d}^{2}\omega^{3}}{3\pi\epsilon_{o}\hbar c^{3}}
\end{equation}
where $p_{d}$,  $\omega$ are  the dipole moment and angular frequency  of the emitter respectively. 

The emitter- cavity coupling rate (g) is given by \cite{Grange2015}
\begin{equation}
g = \cos\theta_{d} \frac{1}{\sqrt{V_{c}}} p_{d} \sqrt{\frac{\omega_{c}}{2\hbar \epsilon_{o}}}
\label{eq16}
\end{equation}
Here $V_{c}$ is the cavity mode volume, $p_{d}$ is the strength of the dipole moment of the emitter, $\theta_{d}$ is the angle between the unit polarization vector of the emitter $e_{d}$ and the cavity field vector $e_{c}$. 
The cavity decay rate $\kappa$ is given by 
\begin{equation}
\kappa= \omega_{c}/Q
\label{eq17}
\end{equation}
where $\omega_{c}$ is the cavity resonant frequency and Q is the cavity quality factor. 
Based on the values for g, $\kappa$, $\gamma$, and $\gamma^{*}$ we can categorize the emitter-cavity coupling in two parts, the strong coupling \& the weak coupling. If 4 |$g$| > |$\kappa - \gamma - \gamma^{*}$|  then the emitter - cavity coupling  is refered to  as  strong coupling and if  4 |g| $<$ |$\kappa$ - $\gamma$ - $\gamma^{*}$ then  it is  case the weak coupling  \cite{Huemmer2013,Shan2013, Gerhardt2019, Press2007}.
For the strong coupling regime the indistinguishability is represented by the following formula \cite{Grange2015}.

\begin{equation}
I=\frac{(\gamma+\kappa)(\gamma+\kappa+\gamma^{*}/2)}{(\gamma+\kappa+\gamma^{*})^{2}}
\label{eq18}
\end{equation}
And in the weak coupling regime, if we have $\kappa$ $>$  $\gamma$+$\gamma^{*}$ then it's the bad cavity regime for which indistinguishability is I$_{BC}$ which is represented by the following formula \cite{Grange2015}.
\begin{equation}
I_{BC}=\frac{\gamma+R}{\gamma+R+\gamma^{*}}
\label{eq19}
\end{equation} 

where R (population transfer rate between emitter and the cavity) is given by
\begin{equation}
R=\frac{4g^{2}}{\gamma+\kappa+\gamma^{*}}
\label{eq21}
\end{equation}
The theoretical extraction efficiency for both the cases is represented by the following formula \cite{Grange2015, Auffeves2010} 
\begin{equation}
\eta=\frac{\kappa R}{(\kappa R)+ (\gamma(\kappa+R))}
\label{eq22}
\end{equation}

\end{document}